\documentclass[aps,prb,epsfig]{revtex4}
\usepackage{graphicx}
\begin{document}
\title{Rescaled density expansions and demixing in hard-sphere binary mixtures}
\author{M. L\'{o}pez de Haro$^\dag$ and C. F. Tejero\footnote{e-mail:
cftejero@fis.ucm.es}}
 \affiliation{Facultad de Ciencias F\'{\i}sicas, Universidad Complutense de
Madrid, E-28040, Madrid, Spain}
\mediumtext
\begin{abstract}
\indent The demixing transition of a  binary fluid mixture of
additive hard spheres is analyzed for different size asymmetries
by starting from the exact low-density expansion of the pressure.
Already within the second virial approximation the fluid separates
into two phases of different composition with a lower consolute
critical point. By successively incorporating the third, fourth,
and fifth virial coefficients, the critical consolute point moves
to higher values of the pressure and to lower values of the
partial number fraction of the large spheres. When the exact
low-density expansion of the pressure is rescaled to higher
densities as in the Percus-Yevick theory, by  adding more exact
virial coefficients a different qualitative movement of the
critical consolute point in the phase diagram is found. It is
argued that the Percus-Yevick factor appearing in many empirical
equations of state for the mixture has a deep influence on the
location of the critical consolute point, so that the resulting
phase diagram for a prescribed equation has to be taken with
caution.

\end{abstract}

\maketitle

Demixing is a common phase transition in fluid mixtures usually
originated on the asymmetry of the interactions between the
different components in the mixture. In the case of  binary
additive hard-sphere mixtures, the existence of demixing has been
studied theoretically since decades, and the issue has been at
times controversial. The importance of this problem resides in the
fact that if fluid-fluid separation occurs, it would represent a
neat example of an entropy-driven phase transition, i.e. a phase
separation based only on the size asymmetry of the spheres. The
solution of the Ornstein-Zernike (OZ) equation with the
Percus-Yevick (PY) closure for additive hard-sphere mixtures by
Lebowitz \cite{Lebowitz1} and the ensuing analysis of the
thermodynamic properties of a binary mixture by Lebowitz and
Rowlinson \cite{Lebowitz2} played a major role in the discussion
and in many of the later theoretical developments. The great
advantage of the PY theory is that it yields analytic expressions
for the equation of state (taking both the virial and the
compressibility routes) as well as for the contact values of the
radial distribution functions. Its main drawbacks are the fact
that it predicts a singularity in the pressure when the total
volume fraction occupied by the spheres is equal to one (which is
physically unattainable) and the thermodynamic consistency
problem, i.e. the fact that the virial and the compressibility
routes lead to different expressions for the equation of state. In
any case, the immediate conclusion that follows from this theory
is that it predicts no separation into two fluid phases. The same
conclusion is reached if one considers the most popular (and
reasonably accurate when compared to simulation data) equation of
state proposed for hard-sphere mixtures, namely the
Boubl\'{\i}k-Mansoori-Carnahan-Starling-Leland (BMCSL)
\cite{Boublik} equation, which can be obtained by an empirical
interpolation of the virial and the compressibility equations in
the PY theory. However, the fluid-fluid segregation found by Biben
and Hansen \cite{Biben} using the OZ equation with the
Rogers-Young closure for highly asymmetric spheres indicated
otherwise and renewed interest in demixing.

One can analyze theoretically the demixing transition in
hard-sphere binary mixtures taking different routes. Of particular
concern to us here is the approach based on the determination of
an equation of state for the mixture beyond the PY theory. Since
no exact result is available, various possibilities arise,
including the solution of the OZ equation with different closures.
However, this would most likely produce no analytical results.
That is the reason behind the many different empirical equations
of state that have been proposed in the literature. Apart from
aiming at providing a reasonably accurate account of the available
simulation data, they attempt to reproduce virial behavior and/or
to comply with consistency conditions of the contact values of the
radial distribution functions (for a recent review of these
consistency conditions see Ref. \onlinecite{Barrio1} and
references therein). In any event, all of them yield the exact
second and third virial coefficients but at the same time inherit
the singularity in the PY theory.

 Coussaert and Baus \cite{Coussaert} have recently proposed an equation of state
for a binary hard-sphere mixture with improved virial behavior
that predicts a fluid-fluid transition at very high pressures
(metastable with respect to freezing). On a different vein,
Regnaut et al. \cite{Regnaut} have examined the connection between
empirical expressions for the contact values of the radial
distribution functions (which amounts to stating the corresponding
equation of state for the mixture) and demixing. In this paper the
demixing transition is analyzed by starting from the exact
low-density expansion of the pressure. Our findings are then
compared with the fluid-fluid separation resulting from different
empirical proposals for the equation of state. It is found that
qualitative differences appear which confirm that demixing in
binary additive hard-sphere mixtures is still an open question
which deserves further investigation.

We consider a  binary fluid mixture of $N=N_1+ N_2$ additive hard
spheres of diameters $\sigma_1$ and $\sigma_2$ ($\sigma_1
>\sigma_2$). The thermodynamic properties of the mixture can be
described in terms of the number density $\rho\equiv N/V$, with
$V$ the volume, the partial number fraction of the big spheres
$x\equiv N_1/N$,  and the parameter $\gamma\equiv
\sigma_2/\sigma_1$ which measures the size asymmetry. For
particles of equal mass, the Helmholtz free-energy  per particle
$f\equiv f(\rho,x,\gamma)$ reads:

\begin{eqnarray}
\label{1} \beta f& = & \ln\left(\rho\Lambda^3\right)-1+ x\ln x +
(1-x)\ln(1-x)\nonumber \\
& + & \sum_{n=1}^{\infty}\, \frac{1}{n}\,B_{n+1}(x,\gamma)\rho^n,
\end{eqnarray}
where $\beta\equiv 1/k_BT$, with $T$ the absolute temperature and
$k_B$ Boltzmann's constant, plays only the role of a scale factor,
$\Lambda$ is the thermal de Broglie wavelength of the particles,
and $B_{n+1}(x,\gamma)$ are the virial coefficients of the
mixture. The pressure $p\equiv p(\rho,x,\gamma)$ is found from the
thermodynamic relation $p=\rho^2(\partial f/\partial\rho)$
yielding:

\begin{equation}
\label{2} \beta p= \rho\left[ 1 + \sum_{n=1}^{\infty}
B_{n+1}(x,\gamma)\rho^{n}\right].
\end{equation}
 By a Legendre transformation we have moreover eliminated from (\ref{2}) the
number density in favor of the pressure  and analyzed the phase
separation using the Gibbs free-energy per particle $g\equiv
g(p,x,\gamma)=f+ p/\rho$.

For fixed $\gamma$, the lower critical consolute point, $p_c$ and
$x_c$, is found from the convexity conditions:

\begin{equation}
\label{4} \left[\frac{\partial^2 g}{\partial
x^2}\right]_{p_c,x_c}=0,\,\,\,\,\,\,\,\,\left[\frac{\partial^3
g}{\partial x^3}\right]_{p_c,x_c}=0.
\end{equation}
Since $g= x \mu_1+ (1-x)\mu_2$, where $\mu_1\equiv
\mu_1(p,x,\gamma)$ and $\mu_2\equiv \mu_2(p,x,\gamma)$ are the
chemical potentials of the components, and from the Gibbs-Duhem
relation $x(\partial\mu_1/\partial x) +
(1-x)(\partial\mu_2/\partial x)=0$, by also fixing the pressure
($p>p_c$), the partial number fractions of the big spheres at
coexistence, $x'$ and $x''$, are obtained from the coexistence
conditions:

\begin{eqnarray}
\label{5}
 & & \nonumber \\
& & \hspace{0.755cm}\left[\frac{\partial g}{\partial
x}\right]_{x'} =\left[\frac{\partial g}{\partial
x}\right]_{x''}\nonumber \\ & &\nonumber \\ & &
\left[g-x\frac{\partial g}{\partial x}\right]_{x'} =\left[g-
x\frac{\partial g}{\partial x}\right]_{x''},
\end{eqnarray}
expressing the equality of the chemical potentials in both phases.

 We have truncated the density expansion in Eq. (\ref{2}) at fifth order since
analytical expressions are known\cite{Kihara1,Kihara2} for
$B_2(x,\gamma)$  and $B_3(x,\gamma)$, while $B_4(x,\gamma)$ and
$B_5(x,\gamma)$ have been evaluated numerically. \cite{Saija,
Enciso1, Enciso2, Wheatley} Demixing has been investigated for
small $\gamma$-values, say $\gamma= 0.05, 0.1, 0.2, 0.3$ and
$0.4$. For $\gamma=0.1$ we have also included a recent numerical
evaluation\cite{Yu} of $B_6(x,\gamma)$.

 We have first kept the density expansion up to second order, where the Gibbs
free-energy can be obtained analytically,  and found that demixing
occurs within this simple approximation. In Fig. 1 we plot the
resulting phase diagram for different $\gamma$-values. We first
note that the coexistence curves are restricted to a small region
in the phase diagram since, for each $\gamma$, the pressure may
only be increased until  the total volume fraction occupied by the
spheres, $\eta = \pi[x + (1-x)\gamma^3]\rho\sigma_1^3/6$, reaches
for the dense phase the highest possible value. Since these
highest volume fractions are not known, all the curves have been
drawn up to the limiting (unphysical) value $\eta=1$. It is seen
that by decreasing the size asymmetry of the spheres the critical
pressure first decreases and the coexistence region widens. After
the critical pressure reaches a minimum ($\gamma\simeq 0.2$) the
trend is reversed. In contrast, the critical composition $x_c$
always increases with $\gamma$. Note that when going for
$\gamma=0.3$ to $\gamma=0.4$ the critical volume fraction $\eta_c$
changes from $0.505$ to $0.819$, i.e. the coexistence region has
shrunk substantially and this explains why by further increasing
$\gamma$ demixing disappears. We point out that this unexpected
qualitative behavior is also encountered by keeping the virial
expansion up to $k$th-order ($k=3, 4, 5$) and similar to the phase
diagram found in Ref. \onlinecite{Coussaert}.

\begin{figure}[tbp]
\includegraphics[width=.70 \columnwidth]{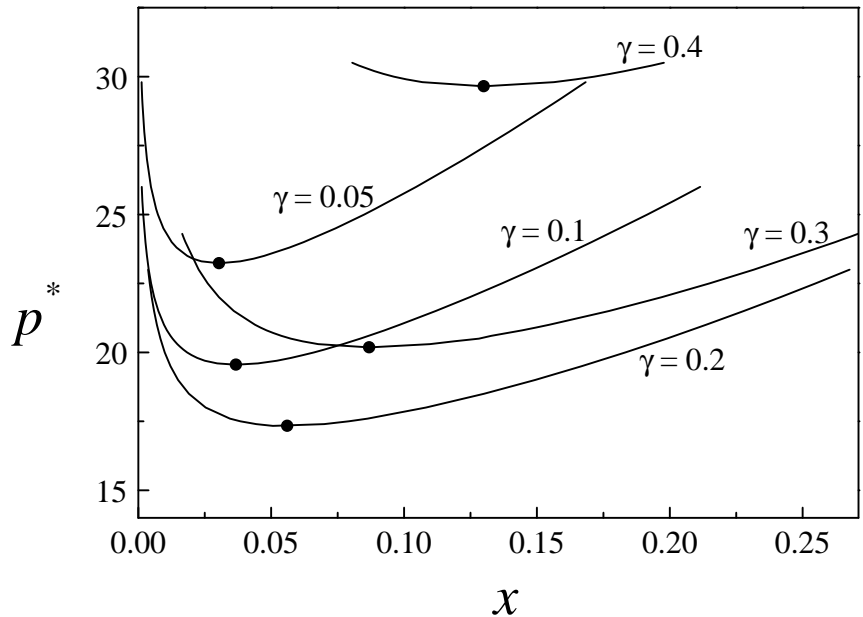}
\caption{
 Phase diagram in the ($x,p^*$)-plane, with $p^*=\beta p\sigma_1^3$, of a binary
additive hard-sphere mixture  for different size asymmetries as
obtained from the second virial approximation. The full dots
denote the critical consolute points while the lines correspond to
the coexistence curves.}
\end{figure}

An important feature for all the reported size asymmetries is that
by  successively incorporating  the third, fourth, and fifth
(sixth) virial coefficients, the critical pressure $p_c$ increases
whereas the critical partial number fraction $x_c$ decreases, as
shown in Table 1.

\begin{table*}
\begin{tabular}{|c|ccc|ccc|ccc|ccc|ccc|}\hline
\multicolumn{1}{|c|}{$$}& \multicolumn{3}{c|}{$B_2$}&
\multicolumn{3}{c|}{$B_3$}& \multicolumn{3}{c|}{$B_4$}&
\multicolumn{3}{c|}{$B_5$}& \multicolumn{3}{c|}{$B_6$}\\ \hline
$\gamma$ & $p_c^*$ & $x_c$ & $\eta_c$
 & $p_c^*$ & $x_c$ & $\eta_c$  & $p_c^*$ & $x_c$ &
$\eta_c$ & $p_c^*$ & $x_c$ & $\eta_c$ & $p_c^*$ & $x_c$ & $\eta_c$
\\ \hline 0.05 & 23.23  & 0.0304 & 0.276 & 50.84 & 0.0294 & 0.428
& 95.18 & 0.0254 & 0.530 & 154.4 & 0.0223 & 0.598 &
 &  & \\
0.1 & 19.55 & 0.0368 & 0.274 &  38.68 & 0.0365 & 0.394 & 67.16 &
0.0333 & 0.476 & 102.6 & 0.0308 & 0.534 &
144.1 & 0.0289 & 0.576\\
0.2 & 17.33 & 0.0561 & 0.339 & 28.48 & 0.0525 & 0.381 & 44.60 &
0.0498 & 0.432 & 64.00 & 0.0489 & 0.476 &
 & & \\
0.3 & 20.18 & 0.0869 & 0.505 & 27.58 & 0.0769 & 0.447 & 39.56 &
0.0727 & 0.463 & 54.00 & 0.0724 & 0.492 &
& & \\
0.4 & 29.65 & 0.1300 & 0.819 & 33.40  & 0.1134 & 0.580 & 43.56 &
0.1066 & 0.548 & 56.46 & 0.1054 & 0.554 &
 &  & \\ \hline
\end{tabular}
\caption{Critical constants $p_c^*$, $x_c$, and $\eta_c$ for
different $\gamma$-values as obtained from the low-density
expansion of the pressure by keeping $n-1$ exact virial
coefficients (indicated as $B_n$).}
\end{table*}

\begin{table*}
\begin{tabular}{|c|ccc|ccc|ccc|ccc|}\hline
\multicolumn{1}{|c|}{$$}& \multicolumn{3}{c|}{H-K}&
\multicolumn{3}{c|}{H-CS}& \multicolumn{3}{c|}{eCS-I}&
\multicolumn{3}{c|}{eCS-II}\\ \hline $\gamma$ & $p_c^*$ & $x_c$ &
$\eta_c$
 & $p_c^*$ & $x_c$ & $\eta_c$  & $p_c^*$ & $x_c$ &
$\eta_c$ & $p_c^*$ & $x_c$ & $\eta_c$ \\ \hline 0.05 & 27.68  &
0.1120 & 0.589 & 29.68&  0.1001 & 0.595
& 3599 & 0.0093 & 0.822 & 1096 & 0.0004 & 0.204\\
0.1 & 38.05 & 0.1114 & 0.624 &  42.97 & 0.0968 & 0.636
& 1307 & 0.0203 & 0.757 & 832.0 & 0.0008 & 0.290 \\
0.2 & 74.77 & 0.0244 & 0.389 & 77.98 & 0.0197 & 0.366
& 653.4 & 0.0537 & 0.725 & - &  - &  -\\
0.3 & 121.5 & 0.0263 & 0.471 & 148.1 & 0.0211 & 0.478
& 581.9 & 0.0998 & 0.738 & - &  -&  -\\
0.4 & 337.3 & 0.0263 & 0.620 & 751.2  & 0.0179 & 0.681 & 663.4 &
0.1532 & 0.766 & - & -& - \\ \hline
\end{tabular}
\caption{The same as in Table I as obtained from the equations of
state proposed by Hamad\cite{Hamad} (H-K and H-CS) and the two
extended Carnahan-Starling equations (eCS-I and eCS-II) of Santos
et al.\cite{Santos2}}
\end{table*}

The last finding seems to be in contradiction with the results by
Coussaert and Baus who rescaled the density expansion in Eq.
(\ref{2}) by assuming that the behavior of the pressure for high
densities is the same as the one found in the PY theory, i.e.

\begin{equation}
\label{6} \beta p = \frac{\rho}{(1-\eta)^3}\left[1 +
\sum_{n=1}^{k}\,C_{n+1}(x,\gamma)\rho^n \right],
\end{equation}
where the upper limit $k$ in the sum indicates that Eq. (\ref{6})
guarantees, through $C_{n+1}(x,\gamma)$, that $\beta p$ has $k+1$
exact virial coefficients at low densities. Coussaert and Baus
found that, when going from $k=3$ to $k=4$, the critical consolute
point moves to lower values of $p_c$ and to higher values of
$x_c$. Since the PY factor, $1/(1-\eta)^3$,  strongly increases
the pressure with respect to the pressure at low densities, the
amplitude of the changes induced by adding one more exact virial
coefficient in Eq. (\ref{6}) led  the  authors to conclude that
{\em the movement of the critical consolute point  casts some
doubt on the convergence of the procedure}. Indeed, for a given
$k$ in Eq. (\ref{6}) the PY factor yields $B_n(x,\gamma)\neq 0$
($n\geq k+2$), which are otherwise zero in the exact density
expansion up to $(k+1)th$-order. Since the virial coefficients
$B_n(x,\gamma)$ ($n\geq k+2$) in Eq. (\ref{6}) depend on $x$
through the exact virial coefficients $B_n(x,\gamma)$ ($n\leq
k+1$), it can be concluded that the {\em different} additional
dependences on $x$ for $k=3$ and $k=4$ are responsible for
shifting the critical consolute point to  lower (higher) values
of $p_c$ ($x_c$). Using a recent numerical evaluation of the sixth
virial coefficient\cite{Yu} for $\gamma=0.1$, we have  confirmed
the lack of convergence of this procedure and found that the
critical consolute point moves from $ p_c^*= 5377 $ and $x_c=
0.0008 $ ($k=3$) to $ p_c^* = 2565 $ and $x_c= 0.0088$ ($k=4$) and
to $ p_c^* = 1572 $ and $x_c^* = 0.0020 $ ($k=5$).

By the same argument, it is important to emphasize that the PY
factor appearing in a great variety of empirical equations of
state for a binary hard-sphere fluid mixture, arbitrarily
introduces, as compared to the exact density expansion, additional
$x$-dependences through the virial coefficients $B_n(x,\gamma)\,\,
(n\geq 4)$  if, as a matter of fact, the empirical equations only
preserve the first two exact virial coefficients. The additional
$x$-dependences are also linked to the particular form of the
empirical equation of state. This would explain why very similar
equations of state do predict no demixing  or demixing, and why,
whenever it exists, the location of the critical consolute point
strongly depends on the proposed empirical equation. In order to
illustrate this point, we have analyzed four proposals for the
equation of state that can be cast in the form of Eq. (\ref{6})
with either $k=3$ or $k=4$ and which predict
demixing.\cite{Santos1} They are the proposals by Hamad
\cite{Hamad} (which we will label as H-K and H-CS depending on
whether the Kolafa or the Carnahan-Starling equation for the
monocomponent fluid is used as input) and the two extended
Carnahan-Starling equations (eCS-I and eCS-II) of Santos et al.
\cite{Santos2} All these equations have been introduced as
alternatives to the BMCSL equation of state. It turns out that
these four equations lead to the first two exact virial
coefficients but differ in the predictions for $B_n(x,\gamma)$
($n\geq 4$). Nevertheless they all give good agreement with the
available simulation results for the compressibility factor.
\cite{Baro, Yau, Mali, Cao} The scatter in the values for the
critical constants shown in Table II is so evident that there is
no indication as to whether one should prefer one equation over
the others, if any. For example, in the proposals by Hamad $p_c^*$
is always an increasing function of $\gamma$. In the cases of the
eCS-I and eCS-II, $p_c^*$ does not follow this trend and can be up
to two orders of magnitude higher. Moreover, the  eCS-II does not
predict demixing for $\gamma\geq 0.2$.

In conclusion, we want to stress our main result, namely that the
location of the critical consolute point of a binary hard-sphere
fluid mixture, and hence the determination of the complete phase
diagram  (including the stability of demixing relative to
freezing) have to be taken with caution when using empirical
rescaled density expansions behaving in a PY-type manner. A deeper
assessment is precluded at this stage and must await the
determination of yet higher virial coefficients.

After this paper was originally submitted, and due to a doubt that
arose from the numerical value of the sixth virial coefficient as
reported in Ref. \onlinecite{Yu}, we became aware through private
correspondence with A. J. Masters of a complementary paper by
Vlasov and Masters.\cite{Vlasov} There, the authors provide the
most recent numerical values of the virial coefficients (up to the
sixth) for various values of $\gamma$. For $\gamma= 0.1$, they
also compute the seventh virial coefficient and the critical
consolute points considering different levels of truncation of the
density expansion. Their critical constants are in good agreement
with the data in Table I.

\vspace{0.2cm}

$^\dag$ On sabbatical leave from Centro de Investigaci\'{o}n en
Energ\'{\i}a, U.N.A.M., Temixco, Morelos 62580 (M\'{e}xico);
e-mail: malopez@servidor.unam.mx

The authors acknowledge financial support from DGAPA-UNAM
(M\'{e}xico) (M. L. H.) and from the Ministerio de Ciencia y
Tecnolog\'{\i}a (Spain) Ref: BFM2001-1017-C03-03 (C. F. T.).

\end{document}